\documentclass[prd, twocolumn, superscriptaddress,floatfix, nofootinbib, preprintnumbers]{revtex4-2}
\usepackage[utf8]{inputenc}
\usepackage[colorlinks=true,citecolor=blue]{hyperref}
\usepackage{amssymb}
\usepackage{amsmath}
\usepackage{graphicx}
\usepackage{footmisc}
\usepackage{url}
\usepackage{multirow}
\usepackage{makecell}
\usepackage{hhline}
\usepackage{comment}
\usepackage{nicefrac}

\usepackage{xcolor}
\def\beq{\begin{equation}}
\def\eeq{\end{equation}}
\newcommand{\bea}{\begin{eqnarray}\begin{aligned}}
\newcommand{\eea}{\end{aligned}\end{eqnarray}}
\newcommand{\cathode}{C\textsc{athode}}
\newcommand{\cwola}{CW\textsc{o}L\textsc{a}}
\newcommand{\anode}{A\textsc{node}}

\begin{document}

\title{Classifying Anomalies THrough Outer Density Estimation (CATHODE)}

% \preprint{EFI-20-5}
\preprint{FERMILAB-PUB-21-389-T}

\author{Anna Hallin}
\email{anna.hallin@rutgers.edu}
\affiliation{NHETC, Dept.\ of Physics and Astronomy, Rutgers University, Piscataway, NJ 08854, USA}

\author{Joshua Isaacson}
\email{isaacson@fnal.gov}
\affiliation{Theoretical Physics Department, Fermi National Accelerator Laboratory, Batavia, IL 60510, USA}

\author{Gregor Kasieczka}
\email{gregor.kasieczka@uni-hamburg.de}
\affiliation{Institut f\"{u}r Experimentalphysik, Universit\"{a}t Hamburg, 22761 Hamburg, Germany}

\author{Claudius Krause}
\email{Claudius.Krause@rutgers.edu}
\affiliation{NHETC, Dept.\ of Physics and Astronomy, Rutgers University, Piscataway, NJ 08854, USA}

\author{Benjamin Nachman}
\email{bpnachman@lbl.gov}
\affiliation{Physics Division, Lawrence Berkeley National Laboratory, Berkeley, CA 94720, USA}
\affiliation{Berkeley Institute for Data Science, University of California, Berkeley, CA 94720, USA}

\author{Tobias Quadfasel}
\email{tobias.quadfasel@uni-hamburg.de}
\affiliation{Institut f\"{u}r Experimentalphysik, Universit\"{a}t Hamburg, 22761 Hamburg, Germany}

\author{Matthias Schlaffer}
\email{matthias.schlaffer@etu.unige.ch}
\affiliation{University of Chicago, Chicago, IL 60637, USA}
\affiliation{Département de Physique Nucléaire et Corpusculaire, Université de Genève, Genève; Switzerland.}

\author{David Shih}
\email{shih@physics.rutgers.edu}
\affiliation{NHETC, Dept.\ of Physics and Astronomy, Rutgers University, Piscataway, NJ 08854, USA}

\author{Manuel Sommerhalder}
\email{manuel.sommerhalder@uni-hamburg.de}
\affiliation{Institut f\"{u}r Experimentalphysik, Universit\"{a}t Hamburg, 22761 Hamburg, Germany}

\begin{abstract}
We propose a new model-agnostic search strategy for physics beyond the standard model (BSM) at the LHC, based on a novel application of neural density estimation to anomaly detection.
Our approach, which we call Classifying Anomalies THrough Outer Density Estimation (\cathode{}), assumes the BSM signal is localized in a signal region (defined e.g.\ using invariant mass). By training a conditional density estimator on a collection of additional features outside the signal region, interpolating it into the signal region, and sampling from it, we produce a collection of events that follow the background model. We can then train a classifier to distinguish the data from the events sampled from the background model, thereby approaching the optimal anomaly detector. Using the LHC Olympics R\&D dataset, we demonstrate that \cathode{} nearly saturates the best possible performance, and significantly outperforms other approaches that aim to enhance the bump hunt (\cwola{} Hunting and \anode{}). Finally, we demonstrate that \cathode{} is very robust against correlations between the features and maintains nearly-optimal performance even in this more challenging setting.

\end{abstract}

\maketitle

\section{Introduction}

While there is compelling theoretical and experimental motivation for new physics to be discovered at the Large Hadron Collider (LHC), it is not possible to perform a dedicated search for every conceivable scenario.  The ATLAS~\cite{atlasexoticstwiki,atlassusytwiki,atlashdbspublictwiki}, CMS~\cite{cmsexoticstwiki,cmssusytwiki,cmsb2gtwiki}, and LHCb~\cite{lhcbtwiki} collaborations have extensive search programs for new physics, but there are more models to search for than can be covered by individual analyses.  Even searches for pairs of particles are largely unexplored~\cite{Craig:2016rqv,Kim:2019rhy}, in part because of the theory space priors guiding analysis development.  The lack of discoveries thus far could therefore be because existing searches do not cover the anomalous regions of phase space.  As a result, it is essential to complement the search program with methods that are more model agnostic.

While some traditional searches for physics beyond the standard model (BSM) provide an interpretation with little dependence on a particular signal model, most searches are optimized with a limited set of benchmarks. 
Only a relatively small number of searches are signal model independent from the start, including analyses that focus on single features (e.g.\ bump hunts) and more multivariate searches that compare data with simulation in a large number of signal regions~\cite{sleuth,Abbott:2000fb,Abbott:2000gx,Abbott:2001ke,Aaron:2008aa,Aktas:2004pz,Cranmer:2005zn,Aaltonen:2007dg,Aaltonen:2007ab,Aaltonen:2008vt,CMS-PAS-EXO-14-016,CMS-PAS-EXO-10-021,CMS:2020ohc,Sirunyan:2020jwk}. 

Recent innovations in machine learning have resulted in powerful new techniques for model agnostic searches in high energy physics.\footnote{See Refs.~\cite{DAgnolo:2018cun,Collins:2018epr,Collins:2019jip,DAgnolo:2019vbw,Farina:2018fyg,Heimel:2018mkt,Roy:2019jae,Cerri:2018anq,Blance:2019ibf,Hajer:2018kqm,DeSimone:2018efk,Mullin:2019mmh,1809.02977,Dillon:2019cqt,Andreassen:2020nkr,Nachman:2020lpy,Aguilar-Saavedra:2017rzt,Romao:2019dvs,Romao:2020ojy,knapp2020adversarially,1797846,1800445,Amram:2020ykb,Cheng:2020dal,Khosa:2020qrz,Thaprasop:2020mzp,Alexander:2020mbx,aguilarsaavedra2020mass,1815227,pol2020anomaly,Mikuni:2020qds,vanBeekveld:2020txa,Park:2020pak,Faroughy:2020gas,Stein:2020rou,Kasieczka:2021xcg,Chakravarti:2021svb,Batson:2021agz,Blance:2021gcs,Bortolato:2021zic,Collins:2021nxn,Dillon:2021nxw,Finke:2021sdf,Shih:2021kbt,Atkinson:2021nlt,Kahn:2021drv,Aarrestad:2021oeb,Dorigo:2021iyy,Caron:2021wmq} which are from the Living Review~\cite{Feickert:2021ajf}.} These \textit{anomaly detection} approaches employ a variety of strategies to be broadly sensitive to new physics with varying methods for modeling the standard model background.  In addition to community challenges such as the LHC Olympics~\cite{Kasieczka:2021xcg} and DarkMachines~\cite{Aarrestad:2021oeb}, these tools have also been applied for the first time to collider data by the ATLAS Collaboration~\cite{collaboration2020dijet}.

\begin{figure}[t!]
    \centering
    \framebox{\includegraphics[width=0.475\textwidth]{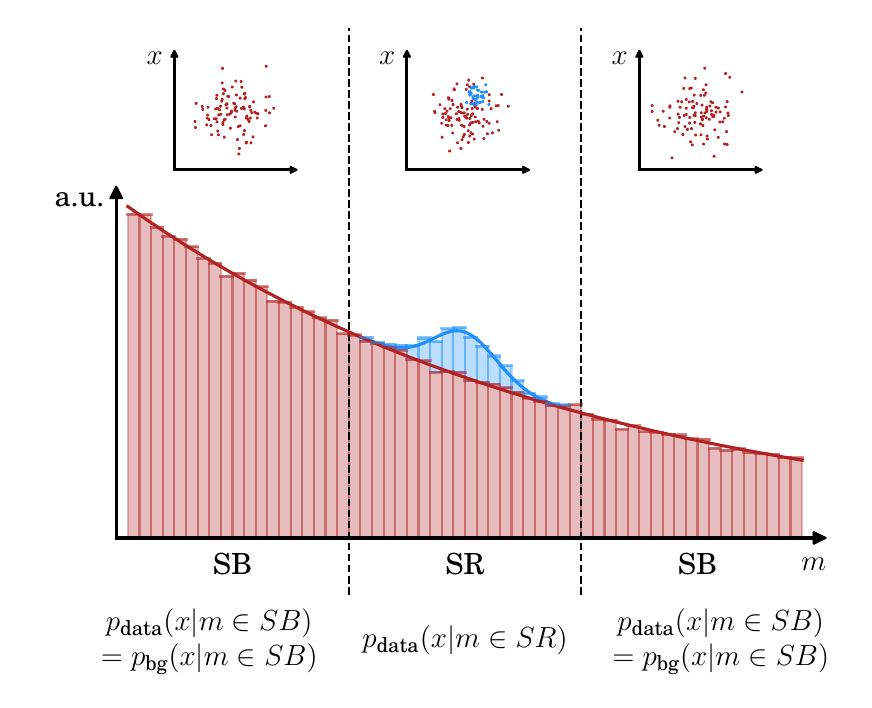}}
    \caption{Schematic view of the bump hunt. The signal (blue) is localized in the signal region (SR). The background (red) is estimated from a sideband region (SB).}
    \label{fig:SB-SR}
\end{figure}

An important class of anomaly detection strategies builds on the bump hunt. The traditional bump hunt assumes that a potential signal is localized in one known feature $m$ (often an invariant mass) and then uses data away from the signal (sideband region or SB) to estimate the background. This setup is sketched in Fig.~\ref{fig:SB-SR}.  The exact location of the signal (signal region or SR) is scanned over $m$.  While broadly sensitive to new physics models with the targeted resonance and nearly independent of simulation for the background modeling, bump hunts are not particularly sensitive to any BSM model.  Machine learning approaches that enhance the bump hunt use features $x$ other than $m$ to automatically amplify the presence of a potential signal. The ultimate goal is to approximate the likelihood ratio between the background and the data in the signal region,
\beq\label{eq:Rofx}
R(x)={p_{\rm data}(x)\over p_{\rm bg}(x)}
\eeq
as this is the optimal test statistic for a data-versus-background hypothesis test~\cite{neyman1933ix}.

Multiple strategies have been proposed for this task.  One approach is based on the Classification Without Labels (\cwola{}) protocol~\cite{Metodiev:2017vrx,Collins:2018epr,Collins:2019jip} in which one trains a classifier to distinguish the SR and SB data.  One of the biggest challenges with the \cwola{} Hunting approach is its high sensitivity to correlations between the features $x$ and $m$. Multiple variations of \cwola{} Hunting have been proposed to circumvent the correlation challenge, such as Simulation Assisted Likelihood-free Anomaly Detection (S\textsc{alad})~\cite{Andreassen:2020nkr} and Simulation-Assisted Decorrelation for Resonant Anomaly Detection (SA-\cwola{})~\cite{1815227}. 

An alternative approach is to learn  the two likelihoods directly   
and then take the ratio.  This is the core idea behind Anomaly Detection with Density Estimation (\anode{})~\cite{Nachman:2020lpy}. The SB is used to estimate $p_\text{bg}(x|m)$ for the background (assuming little signal contamination outside the SR). This likelihood is then interpolated into the SR. Combined with an  estimate of $p_\text{data}(x|m)$ trained in the SR, one can construct an estimate of the likelihood ratio.  The SB interpolation makes \anode{} robust to correlations between $x$ and $m$, although density estimation is inherently more challenging than classification.

In this paper, we propose a new method which combines the best of \cwola{} Hunting and \anode{}. With {\it Classifying Anomalies  THrough Outer Density Estimation} (\cathode{}), we train a density estimator to learn the (usually smooth) background distribution in the SB which we refer to as the ``outer'' region. Then we interpolate it into the SR, but rather than directly constructing the likelihood ratio as in \anode{} (which would require us to also separately learn $p_\text{data}(x|m)$ in the SR), we instead generate {\it sample events} from the trained, interpolated background density estimator. These sample events should follow $p_\text{bg}(x|m)$ in the SR. Finally, we train a classifier (as in \cwola{} Hunting) to distinguish  $p_\text{data}(x|m)$ from $p_\text{bg}(x|m)$ in the SR. 

Using the R\&D dataset~\cite{gregor_kasieczka_2019_4536377} from the LHC Olympics (LHCO)~\cite{Kasieczka:2021xcg}, we will show that \cathode{} achieves a level of performance (as measured by the significance improvement characteristic) that greatly surpasses both \cwola{} Hunting and \anode{}, across a wide range of signal cross sections. \cathode{} easily outperforms \anode{} because it does not have to directly learn $p_\text{data}$ in the SR, and in particular does not have to learn the sharp increase in $p_\text{data}$ where the signal is localized in all of the features. Meanwhile, it outperforms \cwola{} Hunting because of a combination of two effects: one is that in \cathode{}, we can {\it oversample} the outer density estimator, leading to more background events than \cwola{} Hunting has access to (\cwola{} Hunting is limited to the actual data events in the sideband region), and yielding a more powerful classifier. Secondly, the features are slightly correlated with $m$ in the LHCO R\&D dataset, and this slightly degrades the performance of \cwola{} Hunting, while \cathode{} is robust. 

We also compare \cathode{} to a fully supervised classifier (i.e.\ trained on labeled signal and background events) and an ``idealized anomaly detector" (trained on data vs.\ perfectly simulated background). The latter places an upper bound on the performance of any data-vs-background anomaly detection technique, and we show  how \cathode{} essentially saturates its performance. This means that for the first time, a fully-simulation-independent anomaly detection method has been demonstrated to achieve the theoretical upper bound in sensitivity to new physics. The \cathode{} method is basically the best that it could possibly be. 

Finally, as in~\cite{Nachman:2020lpy}, we study the case where $x$ and $m$ are correlated, by adding artificial linear correlations 
to two of the features in $x$. Again we show that \cathode{} (like \anode{}, and unlike \cwola{} Hunting) is largely robust against such correlations, and continues to match the performance of the idealized anomaly detector. 

In this work, we will concern ourselves solely with signal sensitivity, and reserve the problem of background estimation for future study. As long as the \cathode{} classifier does not sculpt features into the invariant mass spectrum, it should be straightforward to combine it with a bump hunt in $m$.

This paper is organized as follows:  Section~\ref{sec:dataset} briefly introduces the LHCO dataset and our treatment of it, and Section~\ref{sec:method} describes the steps of the
\cathode{} approach in detail. Results are given in Section~\ref{sec:results}
and we conclude with Section~\ref{sec:conclusions}. In Appendix~\ref{app:othermethods}, we provide details of the other approaches (\cwola{} Hunting, \anode{}, idealized anomaly detector and fully supervised classifier) considered in this paper. A further study of correlated features is given in Appendix~\ref{app:corrchecks}.

\section{The dataset}
\label{sec:dataset}

For the most part, our treatment of the data (background and signal processes, features, signal and sideband regions) follows \cite{Nachman:2020lpy} closely. Here we will briefly review these choices and also highlight some important differences.

We use QCD dijet events as SM background and $W'\rightarrow X(\rightarrow qq)Y(\rightarrow qq)$ events as signal, where $m_{W'}=3.5~$TeV, $m_X = 500~$GeV and $m_Y = 100~$GeV. These are taken from the original LHCO R\&D dataset~\cite{gregor_kasieczka_2019_4536377}. They are simulated using \texttt{Pythia 8}~\cite{Sjostrand:2006za,Sjostrand:2007gs} and \texttt{Delphes 3.4.1}~\cite{deFavereau:2013fsa,Mertens:2015kba,Selvaggi:2014mya}. The reconstructed particles of each event are clustered into
$R=1$ anti-$k_T$~\cite{Cacciari:2008gp} jets using \texttt{Fastjet}~\cite{Cacciari:2011ma,Cacciari:2005hq}; all events are required to satisfy a single $p_T>1.2$~TeV jet trigger.

\begin{comment}
From these jets the training features
\beq
m_{JJ},\quad m_{J_1},\quad \Delta m_J=m_{J_2}-m_{J_1},\quad \tau_{21}^{J_1}, \quad \tau_{21}^{J_2}
\eeq
are calculated.
Here $J_1$ and $J_2$ refer to the two highest-$p_T$ jets
ordered by their invariant mass (i.e.\ $m_{J_1}<m_{J_2}$), and $\tau_{ij} \equiv \tau_i/\tau_j$ are the $n$-subjettiness ratios~\cite{Thaler:2011gf,Thaler:2010tr}. Finally, $m_{JJ}$ is the invariant mass of $J_1$ and $J_2$ and will be used to define the signal and sideband regions for the enhanced bump hunt. 
\end{comment}

The training features are based on observables constructed by the two highest-$p_T$ jets. The two jets are sorted by their invariant mass, such that $m_{J_1} < m_{J_2}$. The input features used are: the invariant mass of the two jet system ($m_{JJ}$), the invariant mass of the lighter jet, the difference in the invariant masses ($\Delta m_J = m_{J_2} - m_{J_1}$), and the $n$-subjettiness ratios $\tau_{21}^{J_1}$ and $\tau_{21}^{J_2}$. The $n$-subjettiness ratios are defined as $\tau_{ij} \equiv \tau_i/\tau_j$~\cite{Thaler:2011gf,Thaler:2010tr}. 

The signal and sideband regions for the enhanced bump hunt will be defined in terms of the invariant mass of the system: $m_{JJ}\in [3.3,\,\,3.7]$~TeV for the signal region (SR) and its complement $m_{JJ}\notin [3.3,\,\,3.7]$~TeV for the sideband (SB) region. For simplicity, we will specialize to a single $m_{JJ}$ window in this paper, optimally centered on the location of the signal. In practice, as with  any other (enhanced) bump hunt method, one would imagine scanning the SR across the entire $m_{JJ}$ range
and including appropriate trial factors. %Concretely then,  we will take the signal region (SR) to be the dijet invariant mass interval , and the sideband (SB) region to be the complement of the SR (all events with $m_{JJ}\notin[3.3,\,3.7]$~TeV).

%Here we will focus on a SR that is centered on the signal, for simplicity. 

\begin{figure}[t!]
    \centering
    \framebox{\includegraphics[width=0.475\textwidth]{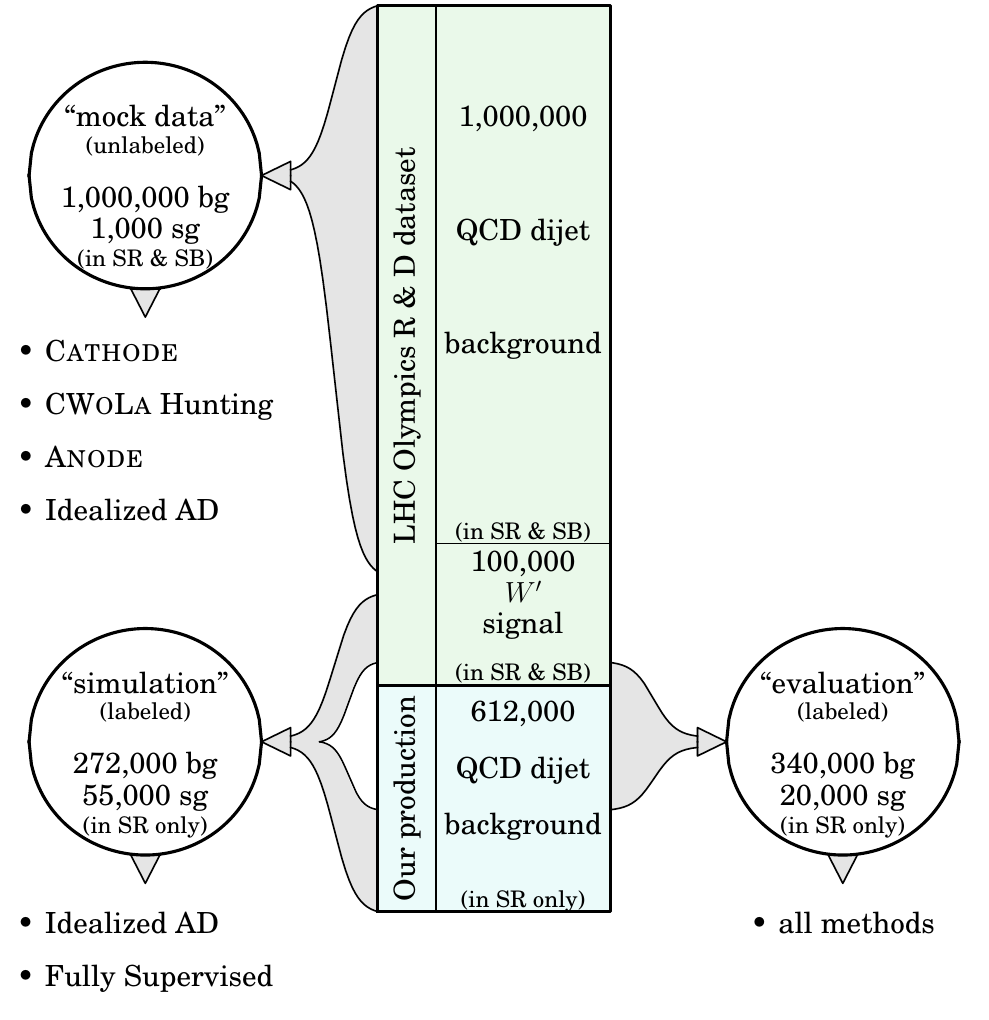}}
    \caption{Visualization of the events and how they are split into datasets. The number of signal (sg) and background (bg) events in each dataset is given. Note that ``simulation'' and ``evaluation'' are only in the SR, so there are some signal events in the SB that are not used at all.}
    \label{fig:data_split}
\end{figure}

In this work we will compare the \cathode{} method against a variety of both simulation-independent anomaly detection (\cwola{} Hunting, \anode{}) and simulation-dependent methods. The simulation-dependent methods will be highly idealized, in the sense that our simulations of background and signal will be assumed to be perfect. Accordingly, we must be very careful about the separation between what we consider as the ``data", i.e. events that would come from an experiment in an actual application of the methods; vs.\ the ``simulation", i.e.~events that would be simulated even in a real world application. Figure~\ref{fig:data_split} visualizes our datasets (see also Table~\ref{tab:datanums} for more
details).

\begin{itemize}

\item For the mock data, we use all of the 1,000,000 SM background events, together with 1,000 (or fewer) signal events, from the original LHCO R\&D dataset. All of the simulation-independent anomaly detection methods  will be trained and validated (model selection) using the mock data alone. 

\item Of the remaining 99,000 signal events in the original LHCO R\&D dataset, approximately 75,000 lie within the SR. For simulation events, we reserved 55,000 of these. For background, we generated an additional 272,000 QCD dijet events specifically in the SR (so with $m_{JJ}\in [3.3,3.7]$~TeV) using the same settings, trigger and data format as the original LHCO R\&D dataset. The fully supervised classifier uses both signal and background simulation events, while the idealized anomaly detector only uses background.

\item Finally we set aside some signal and background events for the common evaluation of all of the methods. These events were not touched during the training or validation of any of the methods. We used the remaining 20,000 SR signal events from the original LHCO R\&D dataset, together with an additionally generated set of 340,000 QCD dijet events in the SR.

\end{itemize}

For our primary benchmark mock dataset (1M background events and 1k signal events), there are 121,352 background events and 772 signal events in the signal region, corresponding to an initial $S/B = 6\times 10^{-3}$ and $S/\sqrt{B} = 2.2$. This is the same benchmark studied in \cite{Nachman:2020lpy} and approximately the same signal vs.\ background composition as Black Box 1 of the LHC Olympics 2020 \cite{Kasieczka:2021xcg}. The purpose of this choice is to ensure that (a) the signal is not too numerous such that a conventional bump hunt in $m_{JJ}$ would already result in a discovery of the signal (obviating the need for any sophisticated anomaly detection method); yet (b) not too few that no anomaly detection method would ever succeed in discovering the signal amongst the background. 

In order to probe this most interesting regime of signal strengths relevant for anomaly detection techniques, we will also perform a scan over different levels of $S/B$ in this work, and we will see the point at which all of the anomaly detection methods fail.

\section{The CATHODE method}
\label{sec:method}

\subsection{Conditional density estimation}

The first step of the \cathode{} method is to train a conditional density estimator on the outer data. Assuming the signal is mostly contained in the SR (as it is here), then the density estimator will learn $p_\text{data}(x|m\notin{\rm SR})\approx p_{\text{bg}}(x|m\notin{\rm SR})$, where $m=m_{JJ}$ and $x=\left(m_{J_1},\,\, \Delta m_J,\,\, \tau_{21}^{J_1}, \,\, \tau_{21}^{J_2}\right)$. 

In this work, we focus on a single baseline density estimator: the Masked Autoregressive Flow (MAF) with affine transformations~\cite{2017arXiv170507057P}. This was used previously in Ref.~\cite{Nachman:2020lpy} and was found to perform well on the LHCO R\&D dataset. (See also Ref.~\cite{Stein:2020rou} for another density estimator that performed well on this dataset.) As in \cite{Nachman:2020lpy}, we will use a base distribution consisting of the unit normal.
In a subsequent publication \cite{ourfuturework} we will compare and contrast different methods for conditional density estimation. For a description of MAFs and normalizing flows more generally, we refer the reader to Refs.~\cite{Nachman:2020lpy,Krause:2021ilc} or to reviews in the ML literature~\cite{2019arXiv190809257K,2019arXiv191202762P}.

As in Ref.~\cite{Nachman:2020lpy}, the features are shifted and scaled to the range $x\in (0,1)$, logit transformed,\footnote{$\mathrm{logit}(x) = \mathrm{ln}\left( \frac{x}{1-x} \right)$}
and finally standardized by subtracting the mean and dividing by the standard deviation of the training set before being passed to the density estimator. This transformation was chosen since it improves the accuracy of the density estimator by turning regions of difficulty (typically sharp edges) into smooth tails, which are easier to learn. 

The mock data in the SB region is split into a training set consisting of 500,000 events, and a validation set consisting of the remaining SB events in the mock data (378,876 to be precise). The validation set is reserved for model selection.

The MAF density estimator\footnote{Derived from the implementation of \url{https://github.com/ikostrikov/pytorch-flows}} 
is trained using PyTorch~\cite{NEURIPS2019_9015} in the SB region for 100 epochs with the Adam optimizer~\cite{kingma2014adam}, a learning rate $10^{-4}$, batch size 256, and batch normalization with a momentum of 1.0. It consists of 15 MADE blocks, with each block consisting of one hidden layer of 128 nodes.  
 This is the same configuration as used in~\cite{Nachman:2020lpy}. The training loss and validation loss are tracked throughout training for each epoch. The ten epochs (model states) with the lowest validation loss are selected for the next step of the \cathode{} method (interpolation and sampling). Since the global minima are used, these 10 epochs do not need to be consecutive. 

\begin{figure}[t!]
    \centering
    \includegraphics[width=0.49\textwidth]{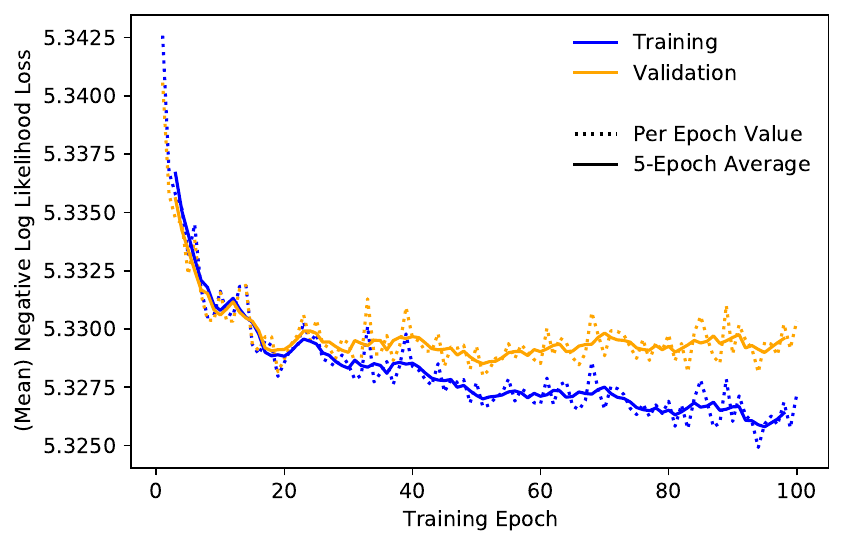}
    \caption{Training and validation loss for the MAF (dotted lines) and the 5 epoch moving average (solid lines)}
    \label{fig:maf_loss}
\end{figure}

The loss curves for one such MAF training are shown as dotted lines in Fig. \ref{fig:maf_loss}, with the moving averages of 5 epochs in solid lines. 

\subsection{Interpolation and sampling}
\label{subsec:interpolation}

The next step of the \cathode{} method is to interpolate the conditional density estimator trained on the SB region into the SR and then sample events from it\footnote{See Ref.~\cite{Lin:2019htn} for another ML-based template method.}. We now describe this process in more detail.

Exactly the same as in the \anode{} method~\cite{Nachman:2020lpy}, this interpolation is automatically handled by the MAF. While the MAF was trained on events with $m\notin{\rm SR}$ to learn a bijective, invertible map $z=f(x;m)$ between the 4d features $x$ and latent space $z$ following the base distribution (unit normal), this function can be queried for any value of $m$, including $m\in{\rm SR}$. In \anode{}, $f$ was used for density estimation, but here we use its inverse
$x=f^{-1}(z;m)$ to produce samples in $x$ following the background distribution in the SR. 

%That is, the automatic interpolation from SB to SR that was used in \anode{} for density estimation (using $f$ to map $x$ to $z$) works equally well for sampling (using $f^{-1}$ to map $z$ to $x$) since it is the same underlying transformation in both cases.

A sample of $N$ events are generated from each of the 10 chosen model states.
The events are then combined and shuffled into a set of $10N$ sample points.  This ensembling procedure gives a more representative set of samples than a single model would\footnote{See Ref.~\cite{EnsembleGanplify} for the impact of ensembling on generative statistics.}. 
In Sec.~\ref{subsec:oversampling}, we will explore the role that $N$ plays in the quality of the anomaly detection task, and the potential benefits of {\it over}sampling the background model in the SR. 

Since we want the sampled synthetic background data to follow the actual data distribution as closely as possible, when sampling we use a matching set of $10N$ $m$ values drawn from the same distribution as the data. To learn the $m$ distribution of the SR data, we perform a kernel density estimate (KDE) fit to the $m$ values in the training set. The KDE was implemented using the Scikit-learn library~\cite{scikit-learn} with a gaussian kernel and a bandwidth of 0.01. To be fully explicit: every sample we produce proceeds from $f^{-1}(z;m)$ with $z\sim {\mathcal N}(0,1)^4$ and $m\sim p_{KDE}(m)$. 

Since the mock data is logit transformed and standardized before being passed to the density estimator, the sampled events are also produced in this transformed and standardized space. They are brought back to the physical space by applying the inverses of the standardization and logit transform, using the SB model parameters (as these were the parameters used by the density estimator). Note that the physical space here refers to the 4d feature space $x=\left(m_{J_1},\,\, \Delta m_J,\,\, \tau_{21}^{J_1}, \,\, \tau_{21}^{J_2}\right)$ and does not include $m$ by construction; sampling from $m$ occurs through the separate KDE step described above.

\begin{figure}[h!]
    \centering
    \includegraphics[width=0.49\textwidth]{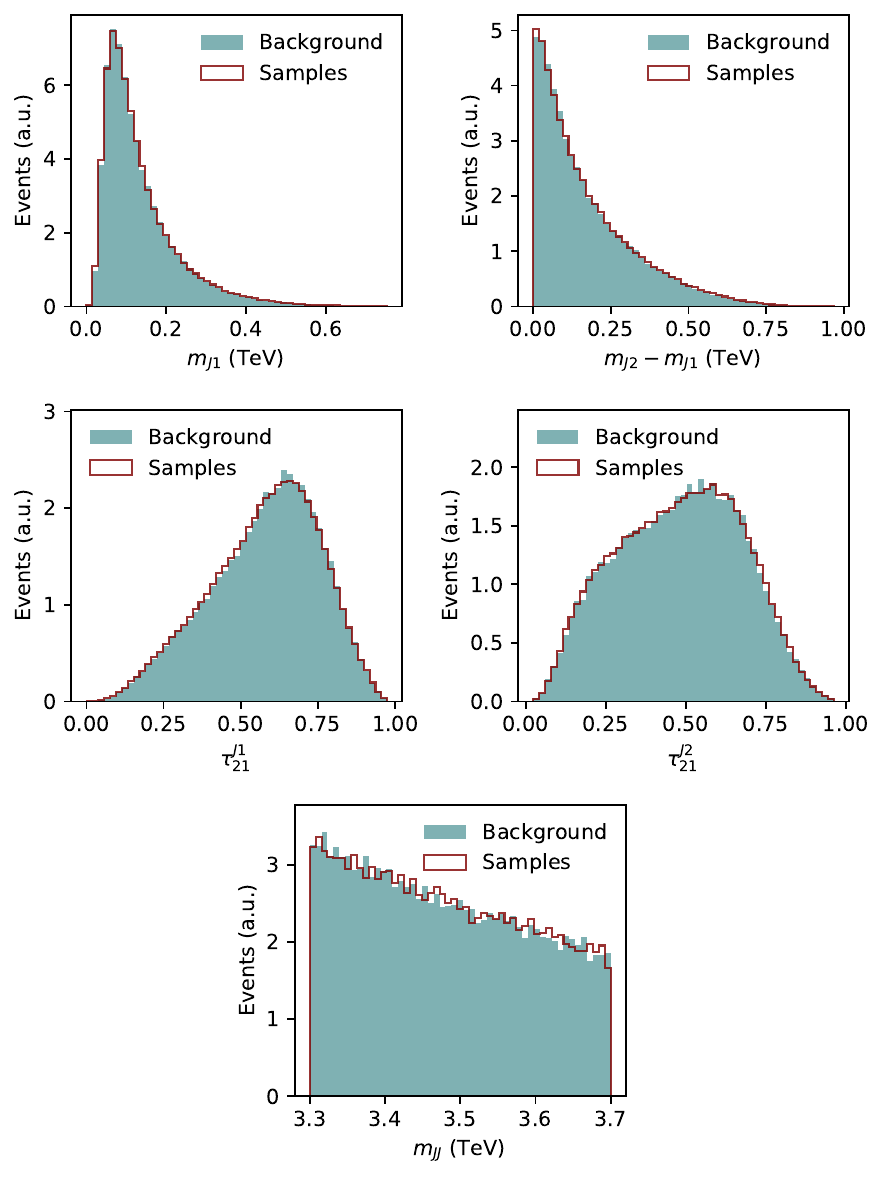}
    \caption{Normalized distributions of the features of the actual background and of the synthetic samples.}
    \label{fig:data_samples}
\end{figure}

The resulting distributions of the sampled events and the mock data background in the validation dataset are shown in Fig.~\ref{fig:data_samples}. One can see that there is a notable overlap between the two distributions in all auxiliary features, as well as on the $m$ distribution drawn from the KDE fit.

\subsection{Classifier}
\label{sec:classifier}

The third step of the \cathode{} method is to train a classifier to distinguish the generated sample events (that should follow the background distribution in the SR) from the mock data (that follow the background plus signal distribution in the SR). For all the variations we will explore (including \cwola{} Hunting), we will use the same classifier architecture. This consists of 3 hidden layers with 64 nodes each and a binary cross-entropy loss.

The binary classifier, %implemented with the Keras \cite{chollet2015keras} API with Tensorflow \cite{tensorflow2015-whitepaper} backend,
also implemented with PyTorch~\cite{NEURIPS2019_9015},
is trained for 100 epochs with a batch size of 128, using the Adam~\cite{kingma2014adam} optimizer with a learning rate of $10^{-3}$. When the classes are imbalanced (as will be the case when we oversample the background model), they are reweighted in the loss computation accordingly,  such that they contribute equally. Note that here classes refer to the sampled events and the mock data, not signal and background events.

For this step, we divide the mock data in the SR in half, reserving 60,000 events for training the classifier and the remaining 60,000 events  for validation (model selection). 
In a real-life application one would want to perform $k$-fold cross validation so as to not throw away half of the events. However, as this is a proof of concept we do not employ this here.

Unless stated otherwise, we sample in total 400,000 events from the MAF generative model (so $N=40,000$ in the description of Section~\ref{subsec:interpolation}), which are distributed equally (200,000 each) into the training and validation set for the classifier. 
Different choices will then be compared in Section~\ref{subsec:oversampling}.

Before the mock data and sampled events are passed on to the classifier, the features are re-standardized, this time using the mean and standard deviation of the SR data features. Here, a logit transformation is not used as it has consistently resulted in sub-optimal anomaly detection performance. 

\begin{figure}[h!]
    \centering
    \includegraphics[width=0.49\textwidth]{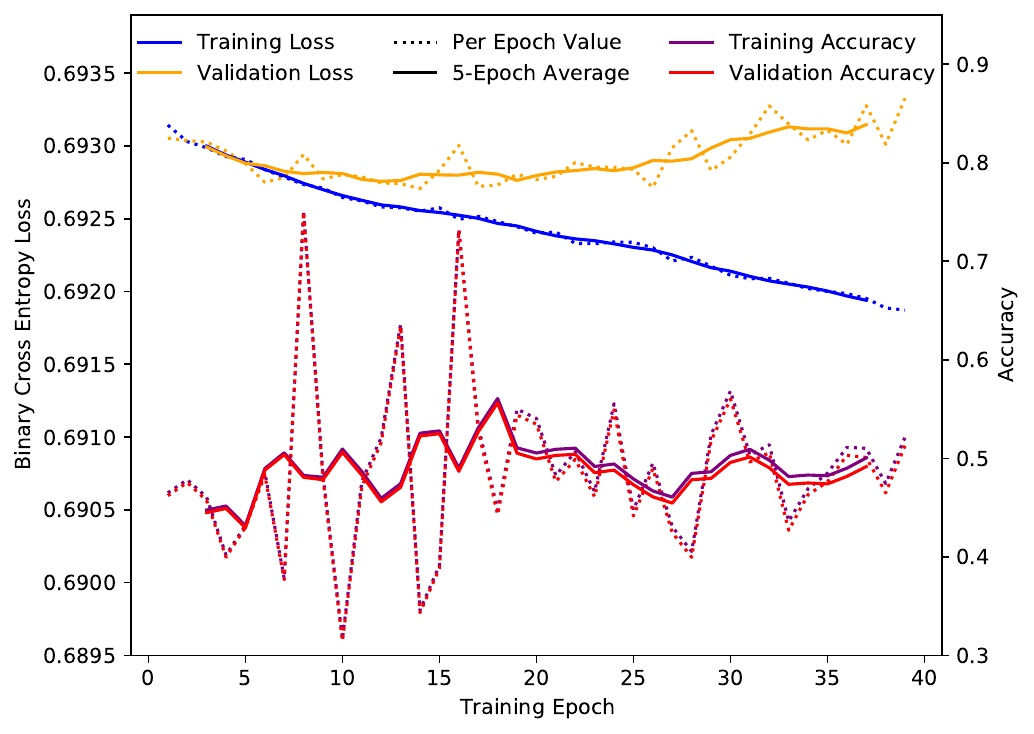}
    \caption{Training and validation loss of the classifier (dotted lines) and the 5 epoch moving average (solid lines) during training. The accuracy is also shown, which in the case of low signal contamination should oscillate around 0.5 if the two classes are indistinguishable.}
    \label{fig:classifier_loss}
\end{figure}

During training, the loss is recorded on the validation set, as shown in Fig.~\ref{fig:classifier_loss}. 
The model states of the 10 epochs with the lowest validation losses are used to construct an ensemble prediction. As in the density estimator ensemble, these epochs do not need to be consecutive. 
In the ensembling, the individual predictions of each data point are averaged. Since the loss is defined with respect to labels indicating whether a data point is from mock data or sampled events, this approach does not rely on any truth information pertaining to the anomaly. 

\subsection{Anomaly detection}

The final step of \cathode{} is to apply the trained classifier to the data in the SR. Recall from the discussion in the introduction that the ultimate goal of an optimal anomaly detector is to learn the likelihood ratio $R(x)$ between the data and background, see Eq.~(\ref{eq:Rofx}). In the presence of an anomaly, we will have 
\beq\label{eq:pdata}
p_{\rm data}(x)=f_{\rm bg}p_{\rm bg}(x)+f_{\rm sig}p_{\rm sig}(x)\,,
\eeq
with a $f_{\rm sig}=1-f_{\rm bg}\ll f_{\rm bg}$ signal (anomaly) fraction. Although this signal fraction is unknown (along with the form of $p_{\rm sig}(x)$), the likelihood ratio $R(x)=p_{\rm data}(x)/p_{\rm bg}(x)$ is nevertheless monotonic with the signal-to-background likelihood ratio. Therefore, if the \cathode{}~method works, the events that are tagged by the classifier as ``data-like" should be signal enriched, regardless of the signal.

In the following section, we will demonstrate the efficacy of the \cathode{} method on the LHCO R\&D dataset.
\begin{comment}
We will quantify our results using the significance improvement characteristic (SIC), which is $\epsilon_S/\sqrt{\epsilon_B}$ vs.\ $\epsilon_S$, where $\epsilon_S$ and $\epsilon_B$ are the signal and background efficiencies of a cut on the classifier. 
Note that these efficiencies can only be calculated using the underlying truth labels that we have access to in the LHCO R\&D dataset. Therefore the SIC curve is being used to demonstrate that the method could find the signal if it was present in the data. In an actual search, where truth labels are not available, one would have to combine the \cathode{} method with a background estimation procedure (e.g.~sideband interpolation as in the bump hunt) and compute a $p$-value under the background-only hypothesis.
\end{comment}
Our performance metric will be the significance improvement characteristic (SIC). The SIC curve is defined as the signal efficiency ($\epsilon_S$) divided by the square root of the background efficiency ($\epsilon_B$), plotted versus the signal efficiency. The background and signal efficiencies are defined based off of a cut on the classifier score. It is important to note that obtaining the SIC curve is only possible through the use of the underlying truth labels available in the LHCO R\&D dataset. Thus, this performance metric is only a means to demonstrate the ability to find a signal in the data if it were present. In practice, one would have to calculate the $p$-value under the background-only hypothesis, while selecting events through the use of \cathode{} and a suitable background estimation procedure (e.g.~sideband interpolation as in the bump hunt). 

As described in Section~\ref{sec:dataset}, in order to improve the statistical significance of these efficiencies, we choose to evaluate all methods on a common test set consisting of 340,000 background events and 20,000 signal events in the SR. This test set is reserved from the outset of the analysis and is never used for the training or validation of any of the methods.

\section{Results}

\label{sec:results}

We first present the results of the \cathode{} method on the original LHCO features, and then we examine the effect of additional correlations between the features. 

Besides \cathode{}, we will also include the performance of several other methods: \cwola{} Hunting~\cite{Collins:2018epr,Collins:2019jip}; \anode{}~\cite{Nachman:2020lpy}; an ``idealized" anomaly detector and a fully supervised classifier. For more details of these methods, see the descriptions in the Introduction and in Appendix~\ref{app:othermethods}.
The idealized anomaly detector, being a classifier between the data and a perfectly simulated background model, sets an upper bound on the performance of any weakly-supervised anomaly detection method that attempts to learn the likelihood ratio between data and background events. Meanwhile, the supervised classifier is trained on labeled background vs.\ signal events. This method sets an absolute upper bound on the performance of any search strategy focused on this signal hypothesis. 

\subsection{Performance on the original LHCO R\&D dataset}
\label{subsec:Original_LHCO_features}

\begin{figure*}[t!] 
    \centering
    \includegraphics[width=\textwidth]{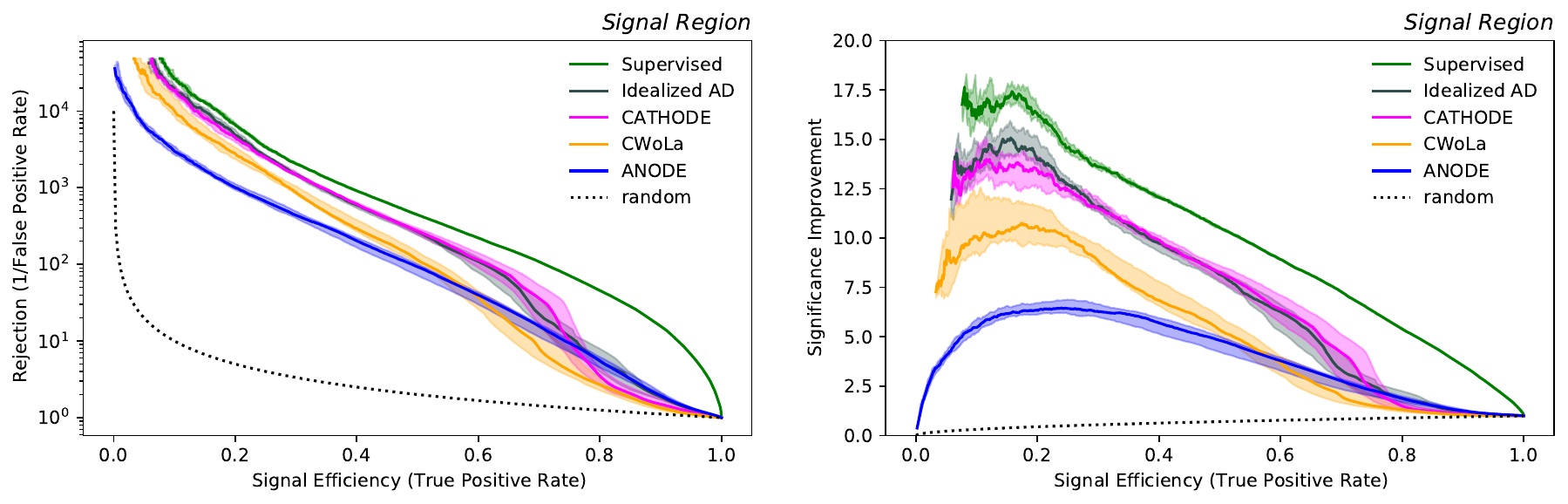}
    \caption{Background rejection (left) and significance improvement (right) of the various anomaly classifiers as a function of the signal efficiency. The solid lines are deduced from a median value of 10 fully independent trainings on the same training, validation and evaluation set. The uncertainty bands quantify the variance from retraining the NNs on the same, fixed dataset and
    are defined such that they contain 68\% of the runs around the median.}
    \label{fig:comparison}
\end{figure*}

\begin{figure*}[hbt!]
    \centering
    \includegraphics[width=\textwidth]{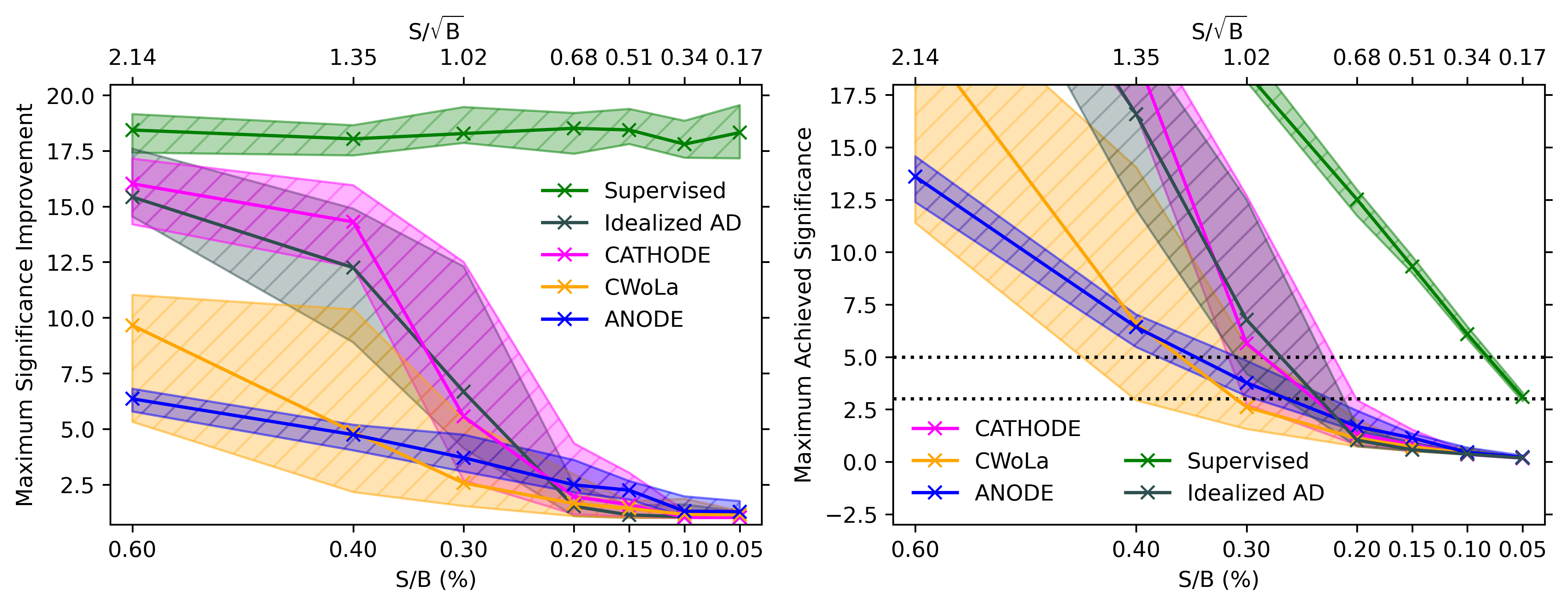}
    \caption{Left: Median maximum significance improvement of each method with 10 different signal injections (leading to a different split of training, validation and evaluation sets in each run) at each decreasing value of signal/background ratios. Here, the 68\% hatched uncertainty bands quantify the
    variance (around the median) from both retrainings of the NN {\it and} random realizations of the training and validation data, including different realizations of the 1,000 injected signal events.  Right: Achieved maximum significance, which is computed by multiplying the uncut significance by the maximum significance improvement. Both plots feature the significance without any cut applied in the upper horizontal axis. The dotted lines on the right hand side denote 3 and 5 $\sigma$ significance values.}
    \label{fig:S_over_B_scan}
\end{figure*}

Figure~\ref{fig:comparison} shows the receiver operating characteristic (ROC) curves and the SIC curves of the different anomaly detection methods trained on our baseline dataset. As described in Section~\ref{sec:dataset}, this consists of 1000 signal events injected into the full background sample, of which 772 are in the SR. 
The curves in Fig.~\ref{fig:comparison} show the median value and 68\% confidence bands of 10 independent trainings, where all steps of each method (e.g.\ both density estimator and classifier for \cathode{}) have been reinitialized in each run. Note that, at this stage, we do not explore the variance due to different realizations of the signal or background events (e.g.\ different choices of the 1000 signal events in the mock data); later in this section, when we explore the performance at smaller $S/B$, the effect of this variation will be included.

We see that overall, \cathode{} outperforms the other weakly supervised methods  across a wide range of signal efficiencies --- a factor of more than 2 compared to \anode{} and a factor of 1.3--2 compared to \cwola{} Hunting. At lower signal efficiencies, \cathode{} reaches a maximum SIC of 14, which represents a significant improvement compared to \anode{}'s 6.5 and \cwola{} Hunting's 11.  A more detailed comparison of \cathode{} with the other methods is as follows:

\begin{itemize}

\begin{comment}
\item \cathode{} outperforms \anode{} presumably for the same reason that \cwola{} Hunting does: instead of trying to model the likelihood function of the data in the SR, including the sharp peak in $x$ where the signal is localized, \cathode{} only has to model the likelihood function of the much smoother background.
\end{comment}
\item Both \cathode{} and \anode{} need to learn the smoothly varying background. However, \anode{} must also learn the sharply peaked distributions in $x$ where the signal is localized (the ``inner" density estimator trained on the SR). This results in a degradation of the \anode{} anomaly detection method and worse performance than \cathode{} and \cwola{} Hunting.

\item As for how \cathode{} is able to outperform \cwola{} Hunting, there are two reasons.
\begin{comment}
The first is that even in the original LHCO R\&D dataset, the features $x$ are slightly correlated (at the percent level) with $m_{JJ}$, 
and this is enough to somewhat degrade the performance of \cwola{} Hunting. The second reason is that 
in \cathode{}, we have the ability to {\it oversample} the background model that we have interpolated from the conditional density estimator trained on the SB events. As described in Sec.~\ref{sec:classifier}, we have trained the \cathode{} classifier on 200,000 events sampled from the background model, whereas \cwola{} Hunting is limited to the background events actually in the sideband region -- which, after dividing 50/50 for training and validation as in \cathode{}, amount to only approximately 65,000 events for training. As we will show in detail below (see Section~\ref{subsec:oversampling}), this oversampling is responsible for a sizable performance boost of \cathode{} over \cwola{} Hunting.
\end{comment}
Firstly, there is a correlation at the percent level between the chosen features in $x$ within the original LHCO R\&D dataset with the search variable ($m_{JJ}$). Since \cwola{} Hunting is very sensitive to correlations, this small correlation is sufficient to degrade the performance compared to that of \cathode{}. Details of the correlation study can be found in Sec.~\ref{sec:shiftedresults}. Secondly, \cwola{} Hunting is limited to only using the events within the sidebands to train the classifier (approximately 65,000 events), while \cathode{} is able to {\it oversample} events from the background model (here 200,000 events are used). These additional events for training allow for a significant performance enhancement of the \cathode{} method. Further details on the effects of oversampling are studied in Sec.~\ref{subsec:oversampling}.

\item Next we turn to the comparison between \cathode{} and the simulation-dependent methods. Recall that the idealized anomaly detector is meant to provide an upper bound on the performance of any data vs.\ background anomaly detection method. Therefore, it is remarkable that \cathode{} achieves essentially the same performance as the idealized anomaly detector. The nearly optimal sensitivity of the \cathode{} method to the signal in the LHCO R\&D dataset indicates that interpolated density estimator is modeling the background in the SR with very high fidelity.

%The fact that \cathode{} is essentially the same as the idealized anomaly detector (they are overlapping within their respective error bands almost everywhere) is truly striking.

 %The fact that the \cathode{} method is nearly saturating it indicates that \cathode{} is achieving close to optimal performance on the LHCO R\&D dataset. Evidently, the background in the SR is being extremely well modeled by the interpolated conditional density estimator.

\item Finally, we see from Fig.~\ref{fig:comparison} that while \cathode{} and the idealized anomaly detector are outperformed by the supervised classifier everywhere (as is to be expected), the difference is larger at higher signal efficiencies. This may be explained by the fact that at higher signal efficiencies, there is simply too much background to find the signal; meanwhile, at lower signal efficiency, the signal is sufficiently localized and the background is sufficiently reduced that the idealized anomaly detector and \cathode{} are more easily able to pick it out.  
%It is worth noting that most jet taggers operate at signal efficiencies (well) below 50\% and so the high signal efficiency region is likely irrelevant. 
% We comment this out since the cross over is not at 50% anymore and also earlier we want to say that CATHODE is up to a factor of 2 better than CWoLa but this is only at higher signal efficiencies. --DS

\end{itemize}

\subsection{Performance at lower signal strengths}
\begin{comment}
So far, our discussion has been focused on the benchmark scenario with 1000 signal events injected into the background sample (corresponding to $S/B\approx 0.6\%$ and $S/\sqrt{B}\approx 2.2$ in the signal region). The performance of the approaches at lower signal rates is explored in Fig.~\ref{fig:S_over_B_scan}. Here, each method is evaluated 10 times for their maximum significance improvement at lower values of the signal/background ratio, each time with a different random separation of signal and background events into training, validation and evaluation sets. 

We see that \cathode{} retains the highest significance improvement among the anomaly detection methods down to a signal fraction of 0.2\,\%, after which point none of the methods can raise the total significance to at least 3 sigma. We also see that across the entire range of relevant $S/B$ values, \cathode{} nearly saturates the upper threshold set by the idealized anomaly detector. The near-optimality of the \cathode{} method is robust against different levels of signal. To say it another way, the sharp degradation in the performance of \cathode{} between $S/B = 0.3\%$ and $0.2\%$ is also seen in the idealized anomaly detector. Therefore it is not due to a failure of the \cathode{} method, but is being set by limited statistics of the data in the signal region. 
\end{comment}

Thus far, the number of signal events injected into the background was fixed at 1000 events ($S/B\approx 0.6\%$ and $S/\sqrt{B}\approx 2.2$). To study the impact of the signal strength in terms of signal improvement, lower signal rates are injected into the background. The injection is done 10 times for each model at each signal rate, and the maximum significance improvement is recorded. Each iteration uses a different random separation into training, validation, and evaluation sets for the signal and background events. The results are shown in Fig.~\ref{fig:S_over_B_scan}.

Above a signal fraction of 0.25\,\%, \cathode{} has the highest significance improvement amongst the different anomaly detection methods. In the region below 0.25\,\%, none of the methods are able to obtain a total significance of at least 3 $\sigma$. We also see that across the entire range of relevant $S/B$ values, \cathode{} saturates the upper threshold set by the idealized anomaly detector. This demonstrates the robustness of the \cathode{} method across a varying level of signal. In particular, the degradation in \cathode{} performance as $S/B$ decreases also occurs for the idealized anomaly detector, so this cannot be attributed to a deficiency in the \cathode{} method.

\subsection{Performance in the presence of correlations}
\label{sec:shiftedresults}

In a realistic application of anomaly detection, the signal and its properties are unknown. Therefore, one needs to be able to choose the set of auxiliary variables $x$ as arbitrarily as possible, in order to gain generic discrimination power through them. However, some anomaly detection algorithms (e.g.\ \cwola{} Hunting) are known to break down once there are significant correlations between $x$ and $m_{JJ}$, thus limiting the choice of candidates for $x$.

\begin{figure*}[t!]
    \centering
    \includegraphics[width=0.49\textwidth, trim={15.7cm 0 0 0},clip]{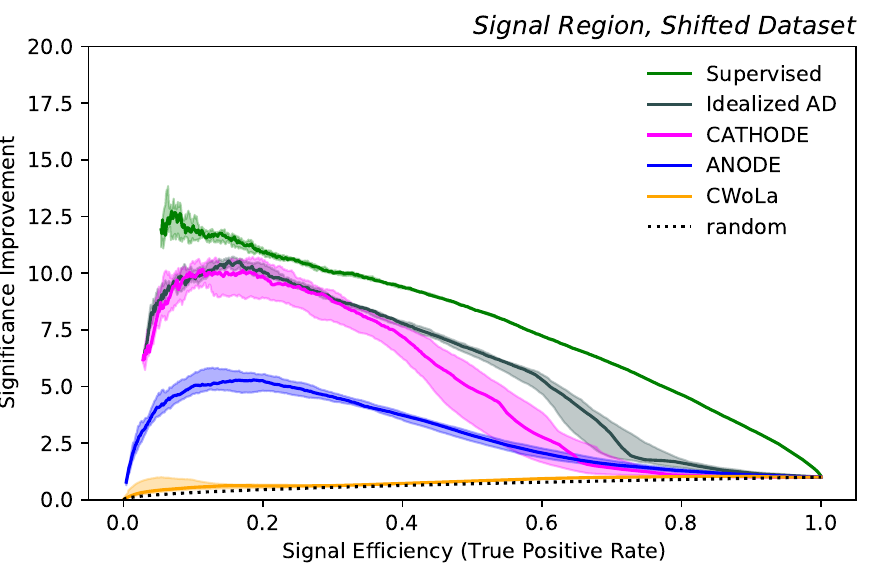}
    \includegraphics[width=0.462\textwidth]{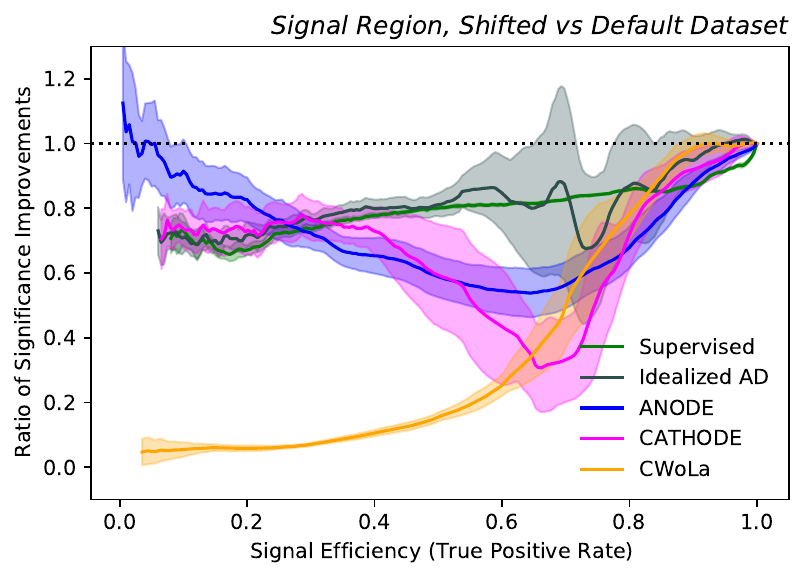}
    \caption{Left: Significance improvement of the various anomaly classifiers as a function of the signal efficiency on the shifted dataset. The solid lines are deduced from a median value of 10 fully independent trainings on the same training, validation and evaluation set. The uncertainty bands are defined the same way as in Fig.~\ref{fig:comparison}. Right: The ratio between the significance improvement with and without the shift on the data applied.}
    \label{fig:comparison_corr}
\end{figure*}

As in~\cite{Nachman:2020lpy}, we test this effect by introducing an artificial correlation between $x$ and $m_{JJ}$ via shifting the features $m_{J_1}$ and $\Delta m$ in each event according to
\bea \label{eq:shift}
	m_{J_1} & \rightarrow m_{J_1} + 0.1 m_{JJ} \\
	\Delta m & \rightarrow \Delta m + 0.1 m_{JJ} 
\eea

The \cathode{} method is applied to the shifted dataset in the otherwise same setup as described in Section~\ref{sec:method}. The same benchmark methods as in Fig.~\ref{fig:comparison} are tested on this shifted data analogously and compared in Fig.~\ref{fig:comparison_corr}.

We see that to varying degrees, each of the different anomaly detection methods (as well as the supervised classifier) suffer from a performance loss due to the shift. In more detail:

\begin{enumerate}

\item Most notably, the \cwola{} Hunting performance breaks down completely. This is completely expected, because the classifier can trivially deduce from the difference in $m_{JJ}$ distribution whether a data point comes from the signal region or sideband, rather than learning the desired likelihood ratio. 

\item Interestingly, the performances of the idealized anomaly detector and the supervised classifier also degrade due to the shift in $x$, with the degradation somewhat larger at lower signal efficiencies. We surmise that this is due to the fact that the classifiers are trained on $x$ alone and not $m_{JJ}$; adding $m_{JJ}$ to $x$ then is effectively like smearing $x$ by another independent random variable. This in turn makes the signal less localized relative to background, which would degrade the performance of even an optimal classifier---especially at lower signal efficiencies where the classifier is benefitting most from the localization of the signal relative to the background.

\item The \anode{} method involves density estimation alone and not the classifier, which means that it does not have the same sensitivities to correlations that \cwola{} Hunting does. However, we see from Fig.~\ref{fig:comparison_corr} (right) that there is a drop in the performance of \anode{} due to the shifted features, primarily at higher signal efficiencies. We attribute this to a combination of a more smeared out and difficult-to-find signal (as in the previous case), as well as worse density estimation in the presence of correlated or noisy features.

\item Finally, we come to the \cathode{} method. Since \cathode{} involves both density estimation and classification, we can think of it as a hybrid of \anode{} and the idealized anomaly detector. 
%Indeed, from Fig.~\ref{fig:comparison_corr} (right), it is striking how \cathode{}'s performance degradation appears to be a ``sum" of that of \anode{} at higher signal efficiencies and the idealized anomaly detector at lower signal efficiencies. 
From Fig.~\ref{fig:comparison_corr} (left), we see that at lower signal efficiencies, \cathode{} is still comparable to the idealized anomaly detector and supervised classifier. Therefore, whatever is degrading the performances of the latter two is also affecting \cathode{} in a similar way. 
Meanwhile, at higher signal efficiencies, \cathode{} is noticeably worse than the idealized anomaly detector and seems to be tracking \anode{} instead. Here we may be seeing the additional effect of poorer density estimation as for \anode{}.

\end{enumerate}

In appendix \ref{app:corrchecks}, we provide further evidence that the classifiers used in \cathode{} and the idealized anomaly detector are suffering from smearing $x$ by the random variable $m_{JJ}$, by adding $m_{JJ}$ to the set of classifier inputs and showing that we more or less recover the lost performance that way.

\subsection{Benefits from oversampling the background model}
\label{subsec:oversampling}

Finally, we turn to a discussion of the benefits of oversampling events from the background model, a unique advantage of the \cathode{} method.
For a more general discussion of the statistical properties of oversampled generative models, see Reference~\cite{2008.06545}.

\begin{figure*}[hbt!]
    \includegraphics[width=0.47\textwidth]{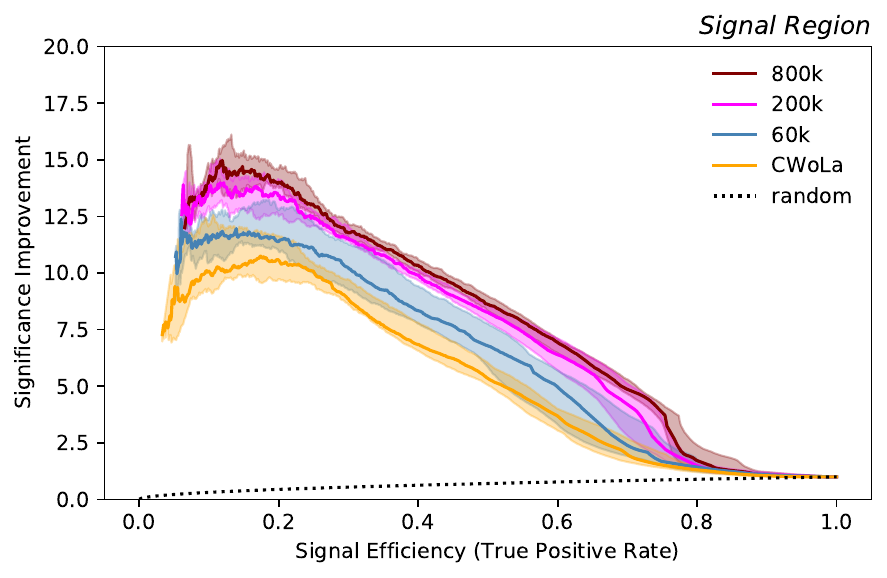}
    \includegraphics[width=0.47\textwidth]{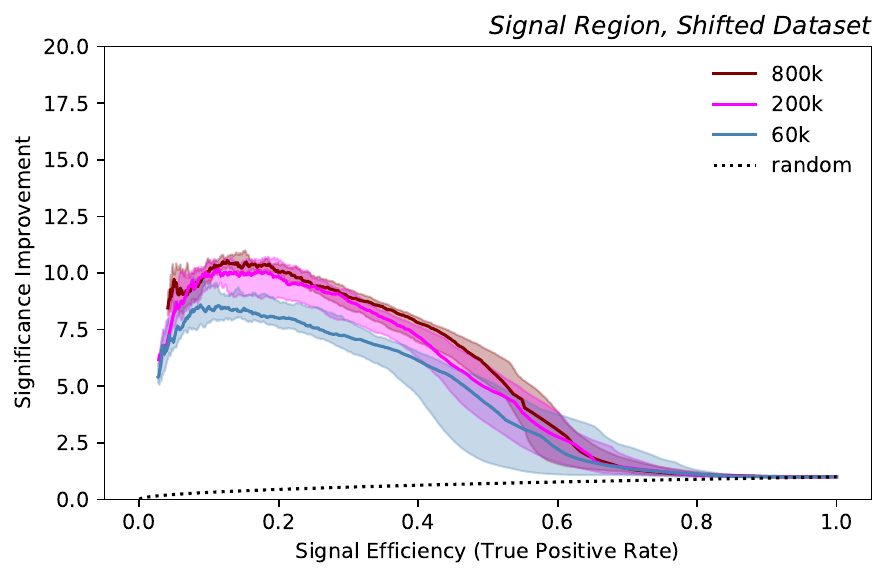}
    \caption{The effect of increasing the number of sampled events when training the classifier. The total number of mock data events in the training set is fixed at 60,000 while the number of sampled events is varied. Left: For the non-shifted data, the performance is boosted by increasing the sample size to 200,000. Increasing the sample size to 800,000 does not provide any further improvement. 
    \cwola{} Hunting, which has access to approximately 65,000 background events in the data,  
    is slightly worse than \cathode{} running on 60,000 samples. Right: For the correlated dataset, increasing the number of sampled events also yields a performance improvement. The solid lines are deduced from a median value of 10 fully independent trainings on the same training, validation and evaluation set. The uncertainty bands in both plots are defined the same way as in Fig.~\ref{fig:comparison}.
    }
    \label{fig:sample_sizes}
\end{figure*}

In Fig.~\ref{fig:sample_sizes} (left), we show the SIC curves for \cathode{} classifiers trained with different numbers of sampled background events, against a baseline \cathode{} classifier trained on 60,000 sampled background events. This baseline is chosen to correspond to the (fixed) number of mock data background events used in the training in the SR. 

As the size of the background sample set is increased from 60,000 to 200,000, the performance improves significantly, especially at lower signal efficiencies. Increasing it further to 800,000 does not provide additional improvement, so we settled on using 200,000 sampled events in the performance plots above.

In Fig.~\ref{fig:sample_sizes} (left), we also include the \cwola{} Hunting's SIC curve for the sake of comparison. We see that even though \cwola{} Hunting was trained with a comparable number (approximately 65,000) of background events in the Short Sideband region (see Appendix~\ref{app:cwola}), its performance is slightly worse than the 60,000 \cathode{} baseline. As discussed in  Section~\ref{subsec:Original_LHCO_features}, this is likely due to small correlations between $x$ and $m_{JJ}$ in the original LHCO R\&D dataset. 

Finally, Fig.~\ref{fig:sample_sizes} (right) shows the impact of varying the sample size for \cathode{} when running on correlated features. Increasing the sample size here yields a modest (but significant) gain in performance.

\subsection{Background estimation}
\label{subsec:bg_estimation}

While the SIC and ROC curves represent useful metrics to assess the performance of different methods, they cannot be used in an actual particle physics experiment, since signal and background labels are not available. Instead, one must combine the anomaly score of \cathode{}, which achieves near-optimal signal sensitivity, with a precise method of background estimation, in order to build a complete search for new physics. 

%See~\cite{Nachman:2020lpy} for a more in depth discussion of these issues. 

In this subsection we will present some preliminary explorations of some background estimation methods that could be combined with \cathode{}. A complete treatment of backgrounds, including the calculation of a calibrated $p$-value, would be well-beyond the scope of this proof-of-concept study; we leave it for future work (or for the actual experimental analyses). 

Probably the most robust way to combine \cathode{} with background estimation would be to perform a ``bump hunt'' and scan several signal region bins that cover the whole $m_{jj}$ mass range. For each of the signal regions, a fit of a parametric background shape to the $m_{jj}$ distribution of events passing a cut on the anomaly score would be performed. Using this fit and its respective uncertainties, one can extract a $p$-value that reflects whether a significant excess is observed.

\begin{figure*}[hbt!]
    \includegraphics[width=0.47\textwidth]{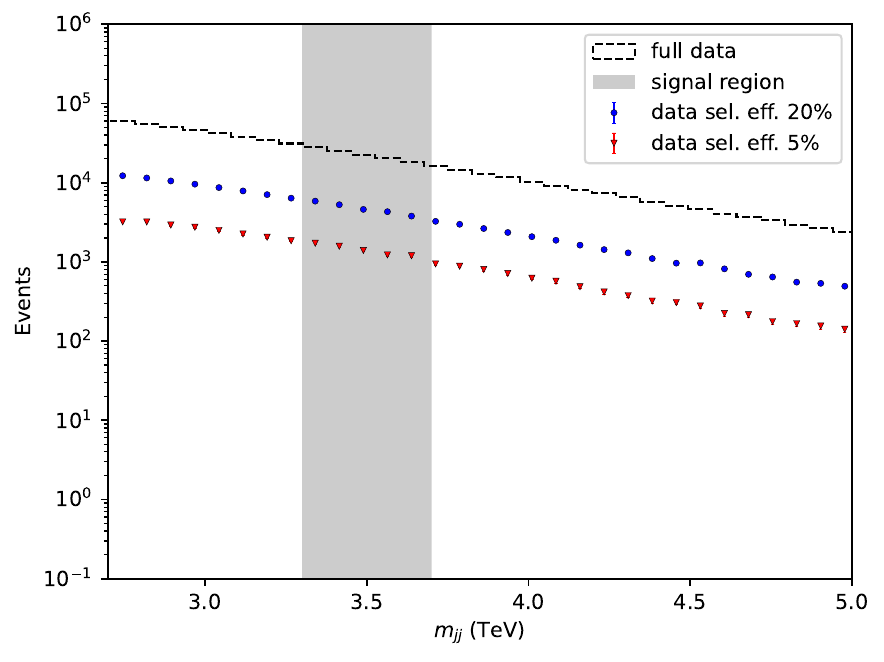}
    \includegraphics[width=0.47\textwidth]{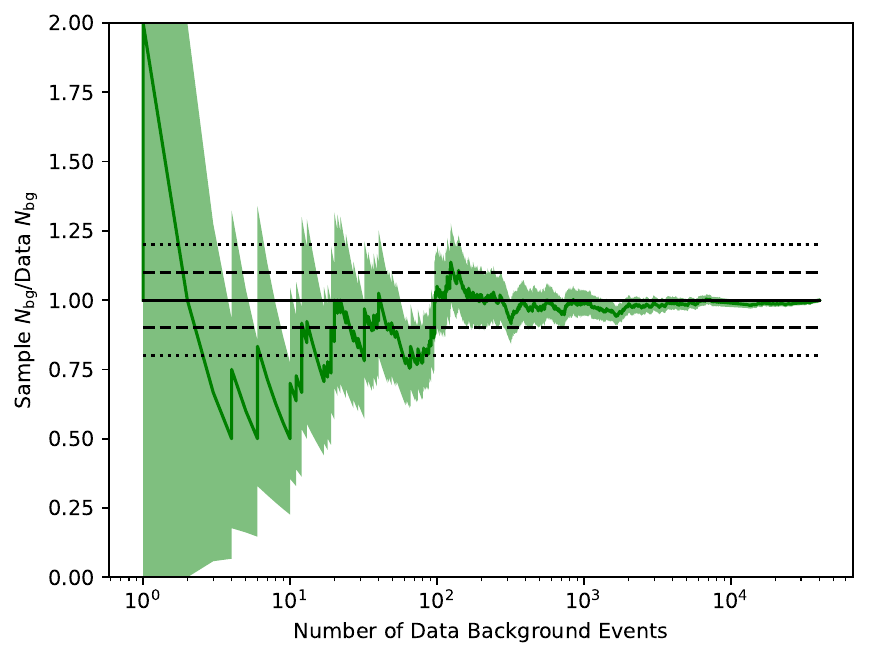}
    \caption{Investigation of background sculpting. Results for training only on background events from the mock data. Left: $m_{jj}$ distributions of background events passing a cut on the classifier output, corresponding to the indicated selection efficiencies. No significant sculpting of features into the background distribution can be observed. Right: Ratio of artificial samples and mock data background events from the signal region passing cut thresholds on the classifier output. There is no significant bias in the background density learned by the density estimator, as the number of events passing a cut on a given threshold is the same (within uncertainties) for background and artificial samples. The uncertainty bands reflect the statistical uncertainty on the number of data and artificial samples, propagated to the ratio.
    }
    \label{fig:bg_estimation}
\end{figure*}

In order for \cathode{} to be used in such a way, it must be able to learn an unbiased estimate of the background density inside the signal region and not sculpt any features into the mass spectrum that could be accidentally found as excesses in a bump hunt. To study this, we trained \cathode{} again on the mock data but this time only using background events and then selected events based on the anomaly score of the model. The anomaly score for background events outside the SR is acquired by simply evaluating these events on the classifier that has been trained to learn the likelihood ratio (up to a transformation) $R(x)=p_{\mathrm{data}}(x)/p_{\mathrm{bg}}(x)$. Since the likelihood ratio only depends on the auxiliary features $x$ and not on $m_{jj}$, this model extends to events from the SBs as well. The dijet invariant mass distributions for the respective selection efficiencies of  20\% and 5\% can be seen in Fig.~\ref{fig:bg_estimation} (left). For reference, the full data distribution is also added. The plot clearly shows that cutting on the \cathode{} model score does not introduce any artificial bumps or features into the $m_{jj}$ distribution and thus it can be used in a bump hunting scenario.

Alternatively, one could imagine another background estimation method where one uses the learned density $p_{bg}(x)$ in the SR to directly estimate the background. Versions of this approach were studied in~\cite{Nachman:2020lpy} (``direct integration'' and ``importance sampling''),  and here we present a simpler and more accurate version of this method: {\it sampling} from $p_{bg}(x)$ in the SR and measuring the background efficiency after a cut on the anomaly score.  If \cathode{} is able to learn an unbiased estimate of the background density in the signal region, a cut on the model output should select as many artificial samples as actual background events. Fig.~\ref{fig:bg_estimation} (right) shows the ratio of the number of artificial samples and background events being selected from these cuts as a function again of the selected background events. This figure illustrates that no significant bias of the model can be observed, since the ratio is around the ideal value of 1 for almost all selection efficiencies and deviations are seen only in regions with low statistics as reflected by the error bands. Comparing with the analogous plot (Fig.~8) in~\cite{Nachman:2020lpy}, we see that \cathode{} presents a much more unbiased background estimate than \anode{}, especially above ${\mathcal O}(10^2)$ background events. This is a reflection of the fact that the likelihood ratio learned by the \cathode{} classifier is much closer to unity on background events than the likelihood ratio constructed in the \anode{} method.

%See for a more in depth discussion of these issues. 

%To investigate a possible bias of the learned density, we used the same model for another study: Using the density estimator, we generated as many artificial samples as background events exist in the signal region. We then evaluated both the artificial samples and background events on the model and cut away events at several fixed thresholds of its output.

\section{Conclusions}

\label{sec:conclusions}

\begin{comment}
We have proposed a new method to detect anomalous regions in data that operates without specific model assumptions beyond the existence of a bump. This new \cathode{} protocol can be seen as a synergy of the \cwola{} Hunting and \anode{} algorithms.
\end{comment}
\cathode{} is a new method for anomaly detection which is model-agnostic beyond the assumption of the existence of a bump. One can think of the \cathode{} protocol as combining the best of the \cwola{} Hunting and \anode{} algorithms.
Similar to these two methods, we first partition data into signal region and sideband according to one feature (typically an invariant mass). As in \anode{}, a conditional density estimator is trained to learn the distribution of sideband data which is assumed to consist purely of background events. This density estimator is then used to generate new background-like events in the signal region. As in \cwola{} Hunting, a classifier is trained to distinguish actual events in the signal region from the background-like events. However, since the background-like events are first transported via the conditional density estimator into the signal region, the result is (as opposed to the result of \cwola{} Hunting) expected to be  robust against correlations between features used to define the signal region ($m$) and features used to train the classifier ($x$). % (the key weakness of \cwola{} Hunting).

\begin{comment}
Using the LHCO R\&D dataset, we empirically test this procedure. Here, \cathode{} achieves superior performance, as measured by the significance improvement characteristic, compared to \cwola{} Hunting and \anode{}. \cathode{} reaches a maximal SIC of 14, compared to 8.5 for \cwola{} Hunting and up to 6.5 for \anode{}. Interestingly, \cathode{} closely approximates an idealized anomaly detector's performance.
%and---for low signal efficiency---even approaches a fully supervised classifier. 
The overall ability of anomaly detection algorithms decreases with the amount of signal present, and \cathode{} achieves a significance of at least 3 sigma down to an S/B ratio in the signal region of about 0.2\%.
\end{comment}
As a benchmark, the LHCO R\&D dataset is used to compare the different anomaly detection algorithms. We find that the \cathode{} method obtains near optimal performance as defined by the idealized anomaly detector.
%, and at low signal efficiency approaches the fully supervised classifier. 
This performance is significantly better than the previous methods of \cwola{} Hunting and \anode{}. In our test point of $S/B = 0.6\%$, \cathode{} has a maximal SIC of 14, while \cwola{} Hunting peaks at 11 and \anode{} at 6.5. While all anomaly detection algorithms degrade as $S/B$ decreases, \cathode{} is able to achieve a significance of at least $3\,\sigma$ until $S/B\approx 0.25\%$.
%, while \cwola{} Hunting and \anode{} are only able to achieve this significance until$S/B\approx 0.3\%$}.

While only one new physics model was used to benchmark the anomaly detection methods, we expect good generalisation of CATHODE to other resonances as the construction only relies on the quality of background estimation from the sideband-regions.

%This gain likely comes from the two main innovations in the construction: using one density estimator instead of two (as in \anode{}) simplifies the learning task, and only comparing events inside (either real or transported to) the signal region removes the problem of correlations for the classification task. \ji{I do not know how to rework this.}

We also explicitly verified that \cathode{} is less sensitive to correlations than other approaches. When artificially increasing the correlation between input features and the mass variable, the \cathode{} performance decreases to a maximum SIC of around 10 while \cwola{} Hunting completely loses discrimination power. However, the artificially increased correlation entangles several issues, including potential information loss and a more difficult task for the density estimator and therefore overestimates the impact of correlation effects.  The enhanced performance in the presence of correlations and the ability to oversample lead to the overall gains from \cathode{} relative to other methods.

\begin{comment}
Given the widespread experimental interest in deploying anomaly detection methods, the increased significance improvement and robustness of \cathode{} compared to previous approaches should directly translate into more sensitive searches. 
\end{comment}
Robust model-agnostic anomaly detection methods are of particular experimental interest. The improvements of \cathode{} over previous approaches should directly translate into more sensitive searches.

\section*{Code and Data}

The code for this paper can be found at \url{https://github.com/HEPML-AnomalyDetection/CATHODE}. The LHC Olympics R\&D dataset can be found at \url{https://zenodo.org/record/4287846}.

\begin{acknowledgments}

The work of AH, CK and DS was supported by DOE grant DOE-SC0010008. The work of BN was supported by the Department of Energy, Office of Science under contract number DE-AC02-05CH11231. GK, TQ, and MSo acknowledge the support of the Deutsche Forschungsgemeinschaft (DFG, German Re\-search Foundation) 
under Germany’s Excellence Strategy – EXC 2121  ``Quantum Universe" – 390833306. This manuscript has been authored by Fermi Research Alliance, LLC under Contract No. DE-AC02-07CH11359 with the U.S. Department of Energy, Office of Science, Office of High Energy Physics. The work of MSc was supported by the Alexander von Humboldt Foundation.
This research was supported in part by the National Science Foundation under Grant No. NSF PHY-1748958.
 JI, CK, and MSc thank Christina Gao for her contributions in the early phase of this project.

\end{acknowledgments}

\appendix
\section{Other methods}
\label{app:othermethods}

In this appendix, we provide more details about the implementation of the various other anomaly detection methods used in the paper. For a summary of the number of events used in each method, see Table~\ref{tab:datanums}.

\begin{table*}[!ht]
\caption{Numbers of events (rounded to the nearest 1,000) used for training, model selection, and evaluation for each method. All methods are evaluated on the same events. } 
\label{tab:datanums}
\begin{center}
  \begin{tabular}{|c|c|c|c|c|}
    \hline
    Method & Type & Train & Validation (model selection) & Evaluation\\
    \hline 
    \multirow{3}{*}{\cathode{}} & density estimator & 500k SB data & 380k SB data & \multirow{11}{*}{\makecell{340k SR background\\20k SR signal}}\\
    \hhline{|~|-|-|-|~|}
    & \multirow{2}{*}{classifier} & 200k SR background samples & 200k SR background samples &\\
    & & 60k SR data & 60k SR data &\\
\hhline{|-|-|-|-|~|}
    \multirow{2}{*}{\anode{}} & \multirow{2}{*}{density estimator} & 500k SB data & 380k SB data &\\
    & & 60k SR data & 60k SR data &\\
\hhline{|-|-|-|-|~|}
    \multirow{2}{*}{\cwola{} Hunting} & \multirow{2}{*}{classifier}
         & 65k SSB data & 65k SSB data &\\  
    & & 60k SR data & 60k SR data &\\
\hhline{|-|-|-|-|~|}
    \multirow{2}{*}{Idealized AD}  & \multirow{2}{*}{classifier}    & 136k SR background &  136k SR background &\\
    & & 60k SR data & 60k SR data &\\
\hhline{|-|-|-|-|~|}
    \multirow{2}{*}{Fully Supervised} & \multirow{2}{*}{classifier} &  136k SR background & 136k SR background &\\
    &  & 27k SR signal &  27k SR signal &\\
    \hline
  \end{tabular}
\end{center}
\end{table*}

\subsection{Idealized anomaly detector}

We start by describing our implementation of the idealized anomaly detector, since all of the other anomaly detection approaches  considered in this paper are approximations of it.

In the idealized anomaly detector, we train a classifier to distinguish the data in the SR from events taken from a perfectly simulated background model. An optimal classifier will approach the likelihood ratio
\beq\label{eq:idealLR}
R_{\rm ideal}(x)={p_{\rm data}(x)\over p_{\rm bg}(x)}=f_{\rm bg}+f_{\rm sig}{p_{\rm sig}(x)\over p_{\rm bg}(x)}
\eeq
where in the second equality we have used Eq.~(\ref{eq:pdata}). Since this is monotonic with the signal-background likelihood ratio, events with high $R_{\rm ideal}(x)$ will also be more likely to be signal than background.

The classifier for the idealized anomaly detector was built using the same network as the \cathode{} classifier, using the same loss, learning rate, and optimizer. For the ``data" class, we use the same training and validation split as for the other anomaly detection methods (60,000 events in the SR each for training and validation). For the ``background" class, we divide up the 272,000 ``simulation" background events evenly into training and validation sets.

\subsection{CWoLa hunting}
\label{app:cwola}

In the \cwola{} Hunting approach~\cite{Collins:2018epr,Collins:2019jip}, one attempts to approximate the likelihood ratio (\ref{eq:idealLR}) by training a classifier to distinguish the events in the SR from the events in a control region (CR) which are assumed to be all background.
The network learns (a monotonic function of) the likelihood ratio given a set of observables ($x$) as:
\beq
R_{\rm CWoLa}(x) = \frac{p\left(x\vert\text{SR}\right)}{p\left(x\vert\text{CR}\right)}
\eeq
Under the further assumption that the distribution of background events in the CR is the same as that of the SR, we have 
\bea
& p(x|\text{SR})=f_{\text{sig}}p(x|\text{sig})+f_{\text{bg}}p(x|\text{bg})   \\
& p(x|\text{CR})=p(x|\text{bg})   
\eea
and we approach $R_{\rm ideal}$.

To enable comparison to \cathode{}, we trained the \cwola{} Hunting network using the same network architecture, loss, learning rate, and optimizer as the \cathode{} classifer. For the SR we take the same $m_{JJ}$ window as all the other methods, and analogously to previous applications on the same dataset~\cite{Nachman:2020lpy}, only the sidebands within 200~GeV wide strips adjacent to the SR in $m_{JJ}$ are used for the CR. We will refer to the CR for \cwola{} Hunting as the Short Sideband (SSB) to distinguish this from the CR used for \anode{} and \cathode{}. 
 This results in a total number of 130,232 background-like events before splitting them equally into training and validation sets. During training, the upper and lower SSB events are reweighted such that they contribute equally to the training and together they have the same total weight as the SR.

\subsection{ANODE}

The \anode{} approach~\cite{Nachman:2020lpy} uses the same interpolated outer density estimator as \cathode{} for the background model. But unlike \cathode{}, it also trains an ``inner" density estimator on the events in the SR. Then it explicitly constructs the likelihood ratio using the two separate density estimators:
\begin{equation}
    R_{\rm ANODE}(x\vert m\in{\rm SR}) = \frac{p_{\text{inner}}\left(x\vert m\in{\rm SR}\right)}{p_{\text{outer}}\left(x\vert m\in{\rm SR}\right)}.
\end{equation}
If the density estimation and interpolation are successful, then $p_{\rm inner}=p_{\rm data}$ and $p_{\rm outer}=p_{\rm bg}$ and we again approach $R_{\rm ideal}$. 

For comparison to \cathode{}, we trained the \anode{} network using the same MAF architecture as used for \cathode{}, with the same loss, learning rate, and optimizer as well. The split of the mock data into training and validation was 50/50 in both the SR and SB, just as in \cathode{}.

\subsection{Supervised classifier}

Finally, to understand the absolute best possible signal vs.\ background discrimination one could ever hope to obtain, we consider the case of labeled data and train a supervised classifier to distinguish directly between signal and background events. 
The supervised network used here is identical to the \cathode{} classifier network, including the loss, learning rate, and optimizer. We used 55,000 signal events and 272,000 background events in the SR, split equally for training and validation.

\section{Adding $m_{JJ}$ as a classifier input}
\label{app:corrchecks}

In the main body of the paper, the approaches all use the features $x$ for data vs.\ background classification, but they do not use $m_{JJ}$. In this appendix, we will explore the effect of adding $m_{JJ}$ as a classifier input in \cathode{}, the idealized anomaly detector, and the fully supervised classifier. 

\iffalse

For the unshifted data, $m_{JJ}$ does not carry much discrimination power between signal and background, and in some sense adding it to the list of classifier inputs is equivalent to adding a random noise variable. In the ideal case, one might expect the classifier can learn to shut off the random input, but in practice, given limited training data and low $S/B$, adding random noise to the anomaly detection classifier can degrade the performance.

For the shifted data, we observed in Fig.~\ref{fig:comparison_corr} that not only \cathode{} but also the idealized anomaly detector and even the fully supervised classifier are worse at separating signal from background when trained on the shifted features. We attributed this to the fact that shifting $x$ by $m_{JJ}$ smears out the signal relative to the background, making it less localized in $x$. Here adding $m_{JJ}$ to the input features can carry a significant benefit, if the classifier can learn to undo the shift. 
\fi

\begin{figure*}[hbt!]
    \centering
    \includegraphics[width=0.49\textwidth, trim={15.7cm 0 0 0},clip]{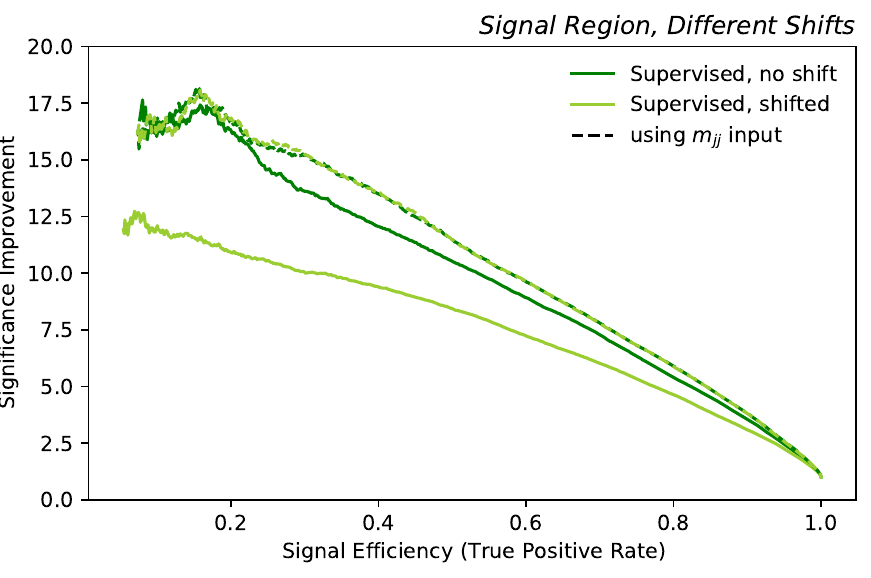}
    \includegraphics[width=0.49\textwidth, trim={15.7cm 0 0 0},clip]{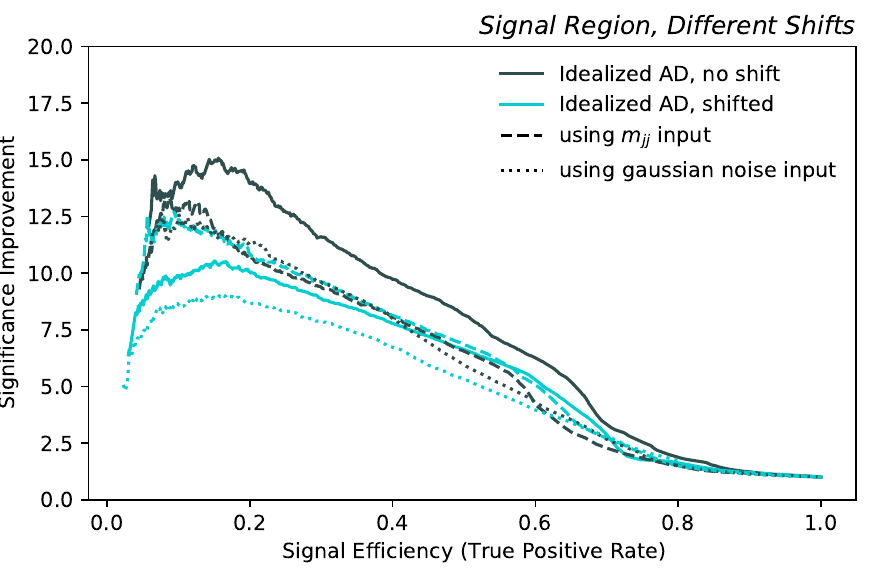}
    \caption{Median significance improvement, deduced from 10 fully independent trainings on the same training, validation and evaluation set, of a supervised training (left) and the idealized anomaly detector (right) in various configurations: using the LHCO R\&D dataset in its default and shifted form, both with and without $m_{JJ}$ as an additional classifier input feature. Moreover, an evaluation on an additional gaussian noise input feature to the classifier is shown for comparison.}
    \label{fig:correlation_cross_comparison}
\end{figure*}

Fig.~\ref{fig:correlation_cross_comparison} (left) illustrates the effect of adding $m_{JJ}$ as an input to the fully supervised classifier trained on the unshifted and the shifted data. We see that the degradation in the fully supervised classifier performance induced by the shifted data is fully recovered by including $m_{JJ}$ as an input feature. This shows that the fully supervised classifier is able to learn to undo the shift of $x$ by $m_{JJ}$. In fact, including $m_{JJ}$ as an input even allows the fully supervised classifier to surpass its original performance, indicating that $m_{JJ}$ does have some (very mild) discriminating ability between signal and background in the signal region. 

The story is a bit more complicated for the idealized anomaly detector. In Fig.~\ref{fig:correlation_cross_comparison} (right), we find that---unlike for the fully supervised classifier---adding $m_{JJ}$ as an input feature to the unshifted data actually {\it degrades} the performance of the idealized anomaly detector (dashed gray). In the signal region, the $m_{JJ}$ distributions are very similar between data and background, plus they are largely uncorrelated with the features $x$. Therefore, including $m_{JJ}$ as an input feature to the idealized anomaly detector is like including the same random noise variable with data and background. In the ideal case, one might expect the classifier can learn to shut off the random input, but in practice, given limited training data and low $S/B$, adding random noise to the anomaly detection classifier can degrade the performance. In Fig.~\ref{fig:correlation_cross_comparison} (right) we confirm the hypothesis that $m_{JJ}$ is like random noise by actually training the classifier with an additional gaussian random noise input instead of $m_{JJ}$ (gray dotted line).  
We find that the degradation in performance is nearly identical to training with $m_{JJ}$.

\iffalse
This is likely due to the fact that $m_{JJ}$ does not have much discriminating power between signal and background {\it in the signal region}, and coupled with the fact that the idealized anomaly detection classifier is trained on data vs.\ background, with such low $S/B$, $m_{JJ}$ is basically just random noise in the signal region. The classifier evidently is not able to learn to shut off the random noise entirely, either because there is insufficient data or the hyperparameters could be further optimized. 
\fi

Meanwhile, the situation is quite different in the shifted case. Here $m_{JJ}$ is no longer functioning like uncorrelated random noise (since $m_{J_1}$ and $\Delta m$ are shifted linearly by $m_{JJ}$). Correspondingly, training with $m_{JJ}$ input does offer some benefit to the idealized anomaly detector (dashed turquoise vs.\ solid turquoise). Interestingly, though, it is more or less capped by the unshifted case (dashed gray). Evidently, the classifier here can learn to undo the shift in $x$, but it still cannot learn to completely shut off the input $m_{JJ}$. 

\begin{figure}[hbt!]
    \centering
    \includegraphics[width=0.49\textwidth, trim={15.7cm 0 0 0},clip]{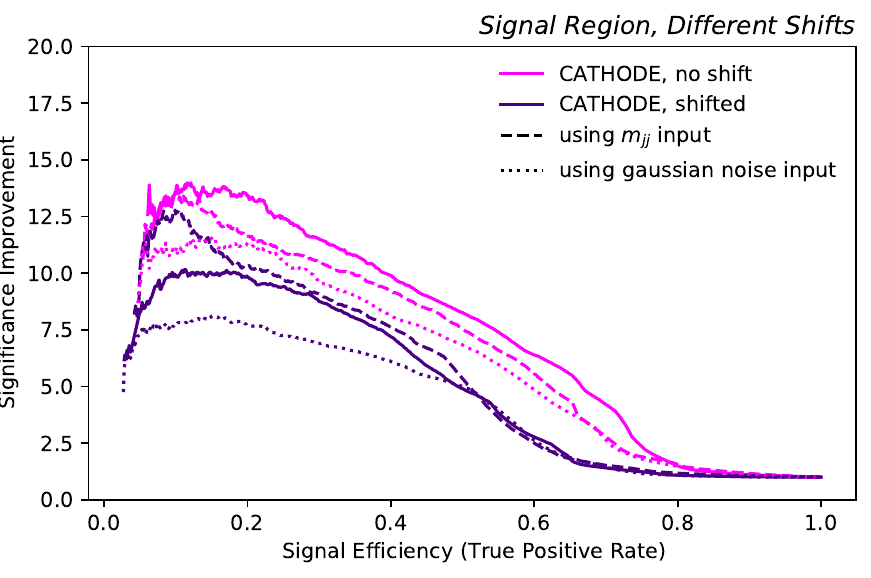}
    \caption{Median significance improvement of CATHODE, deduced from 10 fully independent trainings on the same training, validation and evaluation set, in various configurations: using the LHCO R\&D dataset in its default and shifted form, both with and without $m_{JJ}$ as an additional classifier input feature. Moreover, an evaluation on an additional gaussian noise input feature to the classifier is shown for comparison on the right.}
    \label{fig:correlation_depth_study_CATHODE}
\end{figure}

Finally, in Fig.~\ref{fig:correlation_depth_study_CATHODE} we exhibit the effects of including $m_{JJ}$ in the case of \cathode{}, and we see the behavior is qualitatively very similar to the idealized anomaly detector in nearly all cases. Here the effect of adding gaussian noise input to \cathode{} trained on the shifted data is to degrade the performance even more, whereas adding $m_{JJ}$ input improves the performance, illustrating further how $m_{JJ}$ is not just random noise for the shifted data. 

In summary, we find that adding $m_{JJ}$ to the classifier inputs can have a clear benefit for anomaly detection performance when features are significantly correlated with $m_{JJ}$; however in the absence of correlations adding $m_{JJ}$ can degrade performance if it offers very little discriminating power. Clearly, the issue of feature selection for data vs.\ background anomaly detection is a very important and possibly delicate one, and the story is far from settled on this front. 

\bibliography{HEPML,other}

\end{document}